\documentstyle[12pt]{article}

\def\alwaysmath#1{{\ifmmode{#1}\else{$#1$}\fi}}
\def\he#1{\hbox{\alwaysmath{{}^{#1}}{\rm He}}}
\def\her#1{${{}^#1{\rm He}/{\rm H}}$}
\def\li#1{\hbox{\alwaysmath{{}^{#1}}{\rm Li}}}
\def\msun{${\,M_\odot}$}
\def\hii{H\thinspace{$\scriptstyle{\rm II}$}}
\def\etal{{\it et al.}}
\def\beginapjbib{\begingroup \section*{\large \bf References}
   \parskip=.5ex plus 1.0pt
   \def\bibitem{\par \noindent \hangindent\parindent
      \hangafter=1}}
\def\endapjbib{\par \endgroup}
\begin{document}
\begin{titlepage}
\pagestyle{empty}
\baselineskip=21pt
\rightline{UMN-TH-1305/94}
\rightline{astro-ph/9410058}
\rightline{June 1994}
\vskip .2in
\begin{center}
{\large{\bf What's The Problem With $^3$He?}} \end{center}
\vskip .1in
\begin{center}

 Keith A. Olive

{\it School of Physics and Astronomy, University of Minnesota}

{\it Minneapolis, MN 55455, USA}

Robert T. Rood

{\it Department of Astronomy, University of Virginia, VA 22903}

David N. Schramm, and James Truran

{\it The University of Chicago, Chicago, IL  60637-1433}

{\it NASA/Fermilab Astrophysics Center,
Fermi National Accelerator Laboratory, Batavia, IL  60510-0500}

Elisabeth Vangioni-Flam

{Institut d'Astrophysique de Paris, 98bis
Boulevard Arago, 75014 Paris, France}

\newpage
\end{center}
\vskip .5in
\centerline{ {\bf Abstract} }
\baselineskip=18pt
We consider the galactic evolutionary
history of \he3 in models which deplete deuterium by as much as a factor of
2 to  $\sim$ 15
 from its primordial value
to its present day observed value in the ISM.
We show that when \he3 production in low mass stars (1 -- 3 $M_\odot$)
is included over the history of the galaxy, \he3 is greatly
 over-produced and exceeds
the inferred solar values and the abundances determined in galactic \hii\
regions. Furthermore, the ISM abundances show a disturbing
dispersion which is difficult to understand from the point
of view of standard chemical evolution models.
 In principle, resolution of the problem may lie in either
1) the calculated \he3 production in low mass stars;
2) the observations of the \he3 abundance; or 3) an observational bias
towards regions of depleted \he3. Since \he3 observations in planetary
nebula support the calculated \he3 production in low mass stars, option (1)
is unlikely.
We will argue for option (3) since the \he3 interstellar observations
are indeed made in regions dominated by massive stars in which \he3 is
destroyed.  In conclusion, we note that the problem with \he3 seems
to be galactic and not cosmological.

\noindent
\end{titlepage}
\baselineskip=18pt
\def\la{~\mbox{\raisebox{-.7ex}{$\stackrel{<}{\sim}$}}~}
\def\ga{~\mbox{\raisebox{-.7ex}{$\stackrel{>}{\sim}$}}~}
\def\beq{\begin{equation}}
\def\eeq{\end{equation}}

\section{Introduction}

The utility in an observational determination of a light element
isotope to the theory of big bang nucleosynthesis depends crucially on
our ability to trace the history of that isotope, i.e., to be able to
compare an observed abundance with the prediction of its primordial
value.  Each of the light isotopes presents us with a unique
challenge.  In the case of \he4, we now have a multitude of
observations of \he4 in very low metallicity extragalactic \hii\
regions (Pagel \etal\ 1992; Skillman \etal\ 1994) and because we
expect \he4 to be produced along with oxygen and nitrogen, statistical
analyses allows one to extract the primordial \he4 abundance in a
reasonably straightforward manner (Olive \& Steigman 1994).
\li7 is depleted in stars and is produced in cosmic-ray
nucleosynthesis. It almost certainly has additional sources which
bring primordial values up to observed Pop I values. Standard models,
supported by observational evidence, indicate that the depletion in Pop
II stars and early cosmic ray production are both generally small with
respect to the predicted big bang abundance. Thus the observation of
\li7 in Pop II stars (see e.g., Spite \& Spite 1993) is a good tracer of
the primordial abundance. There are reliable measurements of deuterium
(D or ${}^2$H) in the local interstellar medium (ISM) (Linsky \etal\
1992).  The pre-solar D abundance is determined indirectly by a
comparison between the \he3 abundance in carbonaceous chondrites and
the in gas-rich meteorites, the lunar soil and solar wind (see e.g.,
Geiss 1993). In the former there is a noble gas component with low
\he3 thought to be representative of the true pre-solar \he3
abundance. The latter sample the recent solar wind in which the
initial D has been converted to \he3, so the resulting abundance is the
sum of pre-solar (D + \he3). We know that D is only destroyed in stars
(Epstein, Lattimer \& Schramm, 1976) and the deuterium abundance
should only decrease in time (or remain relatively flat if infall is
dominant).
There may also be some evidence for a measurement of primordial D in a
high redshift, low metallicity quasar absorption system (Songaila
\etal\ 1994; Carswell
\etal\ 1994).  Caution is still warranted with respect to this
observation as it can also be interpreted as a H detection in which
the absorber is displaced in velocity by 80$\,{\rm km\,s^{-1}}$ with
respect to the quasar (see also Vangioni-Flam \& Cass\'{e} 1994;
Steigman 1994; Linsky 1994).  In this context, of all the light
element isotopes of importance to big bang nucleosynthesis, \he3 is
certainly the most difficult isotope to use.  \he3 is both produced
and destroyed in stars and its stellar production/destruction is very
sensitive to the initial mass of the star. The difficulty both in
observing \he3 and in converting the observed quantities to abundances
only compounds the problem in using it as a consistency check on big
bang nucleosynthesis.

In the standard model of big bang nucleosynthesis, there remains only
one key parameter, namely the baryon-to-photon ratio, $\eta$ (Walker
\etal\ 1991).  A comparison between theory and observation for each of
the light elements allows one to set a constraint on $\eta$. Perhaps
the most certain of all of these constraints is the upper limit on
$\eta$ coming from the {\it lowest} observed D abundance in the ISM.
If D is only destroyed then the primordial value must exceed the ISM
value of ${\rm D/H} = 1.65 \times 10^{-5}$ (Linsky \etal\ 1992) and
implies that $\eta_{10}= 10^{10} \eta \la 7$.  (Note that when used in
equations the symbols H, D, \he3, \he4, and \li7 refer to abundances
by number.)  A much tighter constraint is obtained from \he4. Recent
analyses of the \he4 abundance (Olive \& Steigman 1994) indicates that
the $2 \sigma $ upper limit to the \he4 mass fraction is $Y_P < 0.238
(0.243)$ (the larger values allows for a systematic uncertainty). The
corresponding limit on $\eta$ is $\eta_{10} < 2.5 (3.9)$, though as
one can see the upper limit on $\eta$ is very sensitive to the assumed
upper limit on \he4 which in turn is very sensitive to limits placed
on potential systematic errors. The observation of
\li6 in halo stars (Smith \etal\ 1992;
Hobbs \& Thorburn 1994) gives us confidence that \li7 is at most only
slightly depleted (Steigman \etal\ 1993) in these stars.  There is
however, a large systematic uncertainty in the derived \li7 abundance
depending on the assumed model atmospheres. For example, many previous
observations are consistent with
\li7/H $\approx 1.2 \times 10^{-10}$, whereas the recent work of Thorburn
(1993)
finds a systematically higher \li7 abundance, ${\rm \li7/H} \approx 1.9
\times 10^{-10}$. (Given the large numbers of stars observed,
there is almost negligible statistical error in these determinations.)
  Neglecting any depletion or cosmic-ray nucleosynthesis
production, an upper limit of 2 $\times 10^{-10}$ implies that
$1.5 \la \eta_{10} \la 4$. Notice, if we assume that it was deuterium
that has been observed in the quasar absorption system
 at the level of ${\rm D/H} = 1.9 - 2.5 \times 10^{-5}$,
then the value of $\eta_{10}$ is right around 1.5, still consistent with
\li7, and predicts a value of $Y_P \approx 0.23$ in
very good agreement with the
\he4 observations (Cass\'{e} \&  Vangioni-Flam, 1994).  The overall consistency
in the derived ranges for $\eta$ is the chief success of the standard
model of big bang nucleosynthesis.

\section{The Abundance and Chemical Evolution of ${}^3$He}

We now consider the question of \he3. As noted above the solar \he3
abundance is determined from meteorites, the lunar soil and and solar
wind.  There is an increasing body of data on the \he3 abundance
in Galactic \hii\ regions (Balser \etal\ 1994 [BBBRW]). However because of
the great uncertainty in the history of \he3 over the lifetime of the galaxy,
it is very hard to attach a primordial abundance of \he3 in relation
to the observations. Like D, \he3 destruction will be
sensitive to the details of chemical and stellar evolution.
However, in addition,
the models of Iben (1967) and Rood (1972) indicate that
low mass stars, $M \la 2 M_\odot$
are net producers of \he3.  Rood, Steigman and Tinsley (1976)
conjectured that the
\he3\ produced during main sequence hydrogen burning and mixed to the
surface in the first ``dredge-up'' on the lower red giant branch (RGB)
survives the thermal pulsing phase on the asymptotic giant branch
(AGB). The discovery of ``hot bottom burning'' at the base of the
convective envelopes of intermediate mass thermally pulsing AGB stars
(e.g., Renzini \& Voli 1981) raised some concern that \he3 might not
survive. However, recent models of Vassiliadis \& Wood (1993) have
shown little hot bottom burning of \he3 in stars with $M< 5$\msun. For
stars of mass 1--2\msun\ they find the surface \her3\ is $\sim 3
\times 10^{-4}$.  Thus the RGB and AGB winds, and planetary nebulae of
stars $M<2$\msun\ should be substantially enriched in \he3.


  Because of the large input of \he3 rich material into
the ISM from low mass stars Rood \etal\ (1976) argued that the lowest
\he3 abundance observed should serve as an upper limit to the
primordial value and thus set an upper limit for $\eta$.  The argument
yielding a lower limit to $\eta$ based on pre-solar ${\rm D +~^3He}$
was first given in Yang \etal\ (1984), and the argument runs as
follows: First, during pre-main-sequence collapse, essentially all of
the primordial D is converted into \he3.  The
pre-main-sequence produced and primordial \he3 will survive in those
zones of stars in which the temperature is low, $T \la 7 \times 10^6$
K. In these zones
\he3 may even be produced by $p-p$ burning.  At higher temperatures,
(up to $10^8$ K), \he3 is burned to \he4.  If we denote by $g_3$ the
fraction of \he3 that survives stellar processing, then the \he3
abundance at a time $t$ is at least
\beq
\left( {\rm {^3He \over H}} \right)_t \ge g_3
\left( {\rm \frac{D +~^3He}{H}} \right)_p -
g_3 \left( {\rm {D \over H}} \right)_t
\label{yl}
\eeq
The inequality comes about by neglecting any net production of \he3
(and a small amount corresponding to ($1-g_3$) times the fraction
of \he3 that never went into a star).
Of course, Eq. (\ref{yl}) can be rewritten as an
upper limit on {\rm (D +
\he3)/H} in terms
of the observed pre-solar abundances ($t = \odot$) and $g_3$.

In almost all subsequent work, the net production of \he3 has been
neglected.  Values of $g_3$ have been taken to be $\le 1$.  In Yang
\etal\ (1984), an ``extreme" value of $g_3 = 0.25$ was chosen and
combined with the observed pre-solar value of (D + \he3)/H $\le 5.1
\times 10^{-5}$ (Geiss 1993) constrains $\eta_{10} \ga 2.8$. Because
stellar models do not yield values of $g_3$ lower than 0.25, the limit
$\eta_{10} \ga 2.8$ (corresponding to $({\rm D/H})_p \la 8.8 \times
10^{-5}$) should remain intact as a conservative lower bound to
$\eta$.  In Dearborn, Schramm \& Steigman (1986) a more stringent
limit was obtained when values of $g_3$ were integrated over an
initial mass function (IMF). Recently, the question of deuterium
destruction has been examined again. Steigman \& Tosi (1992)
considered several models originally detailed by Tosi (1988) which had
marginal deuterium destruction (by a factor of about 2 total). In
Vangioni-Flam, Olive, \& Prantzos (1994) solar neighborhood models
which destroy deuterium by a total factor of 5 were found, though
values of $g_3$ were required to be somewhat low. The larger depletion
factors found by Vangioni-Flam \etal\ (1994) arise in
part because they employ fewer observational constraints than Steigman
\& Tosi (1992).  In both Steigman \& Tosi (1992) and Vangioni-Flam
\etal\ (1994), \he3 production was ignored.


Here, we show some
results for the evolution of D and \he3 when \he3 production
is included.  We use the estimate for the final surface abundance of \he3
obtained by Iben and Truran (1978). For stars with mass $M < 8 M_\odot$,
\beq
(^3{\rm He/H})_f = 1.8 \times 10^{-4}\left({M_\odot \over M}\right)^2
+ 0.7\left[({\rm D+~^3He)/H}\right]_i
\label{it}
\eeq
where the factor $\left[{\rm (D+~^3He)/H}\right]_i$ accounts for the
premain-sequence conversion of D into \he3 (Yang \etal\, 1984). This
formula probably overestimates \he3 for stars above 5\msun\ because
of the neglect of hot bottom burning but underestimates \he3 for $M <
2$\msun~because of the steeper dependence of stellar lifetime on mass
in that range.
In Figure 1, the differential yield is shown as a function of stellar mass.
Specifically, we plot the mass fraction of \he3 ejected times the IMF
and normalized to the initial mass fraction of D + \he3 corresponding to
(D + \he3)/H $= 9
\times 10^{-5}$.
The \he3 yield was taken from eq. (\ref{it}) for masses $\la 8
M_\odot$ and from Dearborn, Schramm, \& Steigman (1986) for larger
masses, the IMF is a simple power law $\propto m^{-2.7}$ (normalized between
0.4 and 100\msun\ ). The ejected mass is given by
$.89 M - .45 M_\odot$ for $M < 9 M_\odot$ and $M - 1.5 M_\odot$
otherwise (Iben and Tutukov 1984). This figure
clearly shows the importance of the \he3 production in stars with
masses between 1 and 3 $M_\odot$.  Recent work by Tosi (1994) and
Galli \etal\ (1994) also considers the effects of \he3 production.
Stars of various masses contribute differently to the evolution of
\he3.  Massive stars ($> 8M_\odot$) systematically destroy it with
an efficiency increasing with mass.  $g_3$ ranges from 0.3 (at $8M_\odot$)
to 0.11 (at $100 M_\odot$), according to Dearborn Schramm \& Steigman
(1986).  As can be seen from Eq. (\ref{it}),
low mass stars ($M < 3 M_\odot$) are thought to be prolific producers
of \he3 through the $p-p$ chain (Iben \& Truran, 1978), but their yield
is uncertain due to the complexity of the late phases of the stellar
evolution in this mass regime, especially the AGB stage.  Thus, as
in previous work, we have at times taken $g_3$ as a free parameter.
We will let $g_3 =
(x,~y,~z)$ denote the value of $g_3$ at 1,~2,~3\msun.

Vangioni-Flam \etal\ (1994) have explored various combinations of
star formation rates
(SFRs), IMFs, and values of $g_3$ leading to significant D destruction
without overproducing \he3.  All cases required that $g_3$ be less than
1 for $1 < M/M_\odot < 3$ (see their tables 2 and 3).
For example, starting with D/H = $7.5 \times 10^{-5}$ and \he3/H
= $1.5 \times 10^{-5}$, they found that the theoretical evolution can
be made consistent with the observed values provided that $g_3
=0.5,0.5,0.3$ for a simple
star formation rate (SFR) proportional to the mass in gas and a power
law IMF.
The evolution of D and \he3 is followed using a classical closed box
evolutionary model taking into account the delay between star formation
and matter ejection for low mass stars (i.e. the instantaneous
recycling approximation is relaxed).

We can get a good
idea as to the magnitude of the effect on the evolution of \he3 as
$g_3$ is increased to include \he3 production.
To begin with, let us assume an initial value of
$\eta_{10} = 3$, corresponding to a primordial ${\rm D/H} \approx 7.5 \times
10^{-5}$ and ${\rm \he3/H} \approx 1.5 \times 10^{-5}$, as in
the model of Vangioni-Flam \etal\ (1994) above.  When $g_3 =
(0.5,~0.5,~0.3)$,
${\rm (D +
\he3)/H} \simeq 5 \times 10^{-5}$, at the time of the formation
of the solar system, and is acceptable within 2 standard deviations.
When $g_3 =
(1.0,~0.7,~0.7)$, ${\rm (D + \he3)/H} \simeq 6.5 \times 10^{-5}$ or
4.5 standard deviations higher than the solar value.  In Figure 2, we show the
same result (labeled Model 1) when $g_3$ is adopted from Eq.
(\ref{it}).  The corresponding value of $g_3$ is (2.7,~1.2,~0.9).
Clearly there is something wrong.

To test the robustness of this apparent disaster, we have also tried
solar neighborhood models which destroy even more deuterium. If we
assume $\eta_{10} \sim 1.5$ with primordial values of ${\rm D/H}
\approx 2.5 \times 10^{-4}$ and ${\rm \he3/H} \approx 2 \times
10^{-5}$, then the corresponding values of $g_3$ are $g_3 =
(1.5,~0.9,~0.8)$. Note that $g_3$ is lower in this case because the
assumed initial value of (D + \he3)/H is high (cf. eq.(\ref{it})).
What is important however is the product of $g_3$ and $\left[{\rm (D +
\he3)/H}\right]_i$.  To achieve this amount of deuterium destruction,
we have assumed an exponentially decreasing SFR, and the same
power-law IMF (labeled Model 2).  The resulting time evolution is
shown in Figure 3. As one can see from the figure, apart from the
evolution of D (where the model was chosen to destroy D appropriately)
the resulting \he3 (and D + \he3) at the solar epoch and today look
anomalously high compared to the data.  To bring the evolutionary
curves of D/H and (D + \he3)/H into agreement with the data, a value
of $g_3$ no greater than (0.1,~0.1,~0.1) is necessary (Model 2.1).  We
have also taken a larger value of $\eta_{10} \sim 4$ which only
requires a deuterium destruction factor of about 2 (Model 3).  As seen
in figure 4, even though D + \he3 is somewhat acceptable at
$t=t_\odot$, \he3 is still greatly overproduced. Even models with
substantial amounts of infall did not remedy the overproduction
\he3. It appears therefore, that the discrepancy between the chemical
evolution models and the data (taken at face value) is a real effect.

Our results are summarized in the table.  $\sigma$ denotes the gas
mass fraction, D${_o}$ is the present and local interstellar abundance
of deuterium, and $Z$ is the overall metallicity.  As defined above,
models 1, 2, 3 differ by the value of the primordial D/H abundance
(respectively, $7.5 \times 10^{-5}$, $2.5 \times 10^{-4}$ and $3.5
\times 10^{-5}$).  The corresponding values of $g_3$ are
(2.7,~1.2,~0.9), (1.4,~0.9,~0.8), and (4.4,~1.6,~1.1).  Model 2.1 is
similar to that of model 2 except that a $g_3 = (0.1,~0.1,~0.1)$ has
been adopted.  The star formation rates have been adapted in order to
obtain a reasonable amount of D destruction, with an IMF proportional
to $m^{-2.7}$, between 0.4 and 100$\,{M_\odot}$. The star formation
rates we use are: Model 1: ${\rm SFR} = 0.25 \sigma(t)$; Model 2: ${\rm SFR}
=0.67e^{-t/2}$; Model 3: ${\rm SFR} = 0.2 \sigma(t)$.

\section{Discussion}

How can we make any sense of the results of chemical evolution models
in comparison to either the data from the solar system or the galactic
\hii\ regions which show \he3 between $1-5 \times 10^{-5}$?  The first
question we might ask is whether or not stars actually produce \he3.
Indeed, even from the very first observations of \he3, Rood \etal\
(1984) made the suggestion that the build up of \he3 on the main
sequence might be suppressed by non-convective mixing and that the
non-production of \he3 might be correlated with the overproduction of
${}^{13}$C observed in some stars. More recently Hogan (1994) has
suggested that the apparent production of ${}^{13}$C in stars on the
upper RGB suggests a \he3 destruction mechanism. Another suggestion by
Galli \etal\ (1994) is that the ${}^3{\rm He} + {}^3{\rm He}
\rightarrow {}^4{\rm He} + 2p$ reaction has a large low energy
resonance which would greatly reduce the equilibrium abundance of \he3
during $pp$ cycle burning. As seen in the table for Model 2.1, {\em
if} \he3 production in low mass stars can be inhibited {\em and} \he3
destruction at the level of 90\% can be achieved, then the chemical
evolutionary models can be made to fit the data (and if $g_3$ can be
tuned down the lower limit on $\eta$ will be correspondingly reduced).

In contrast, we have observational evidence.  Recently Rood, Bania, \&
Wilson (1992) reported the first detection of \he3 in a planetary
nebula. Further observations reported in Rood \etal\ (1995) show the
detection in NGC 3242 persists over four observing epochs with an
abundance now estimated to be ${\rm \he3/H} \sim 1 \times 10^{-3}$.
There are tentative detections in two other PNe and there is no hint
that the PNe observed are particularly atypical.  In addition, Hartoog
(1979) has observed \he3 in hot horizontal branch stars. While the
observed abundances are generally thought to be strongly affected by
diffusion, they at the least show that some \he3 survives the first
ascent of the RGB (Ostriker \& Schramm, 1994) and are in reasonable
agreement with the stellar evolution models. In conclusion, we would
argue that there is evidence that $g_3$ for solar type stars is large.

If the production factors of \he3 are correct, then why are the
abundances of \he3 in the solar system and in galactic \hii\ regions
so low relative to calculated values?  This is particularly puzzling,
since the stars which produce \he3 do so on relatively long
time-scales.  That is, we would expect \he3 to be well mixed in the
galaxy.  This expectation and a view of the data in galactic \hii\
regions (BBBRW) may in fact already provide a clue to the solution of
the \he3 problem.  The data show a large dispersion of \he3 with
respect to either galactocentric distance, or fraction of ionized
\he4.  Just the fact that there is such a real spread in values is
cause to worry if we believe that \he3 should be well mixed.


If instead the \he3 data is viewed as a function of the mass of the
\hii\ region as in Figure 5 (Balser \etal\ 1995 [BBRW95]), one finds an
interesting and perhaps not unexpected correlation.  The abundance of
\he3 appears to decrease as the mass of the region is increased. The
correlation is real at the 98\% CL with respect to a power-law fit
also shown in Figure 5.  The observed spread in the \he3 concentration
in these regions is significantly greater than the observed spread in
elemental abundances in disk stars at any age (Edvardsson \etal\,
1993).  There are at least 2 ways such a correlation might arise.  The
first comes about in converting the observed line parameter of the
${}^3{\rm He}^+$ hyperfine line to a ${}^3{\rm He}/{\rm H}$ abundance
ratio. Basically the presense of ``structure'' in the form of higher
density subregions will always lead to higher abundances than when the
\hii\ regions are modeled as homogeneous spheres as in BBBRW. The
plotted points include preliminary structure corrections (see BBRW95
for details). The more massive \hii\ regions are on the whole more
distant (for obvious observational reasons). They could have
unresolved ``structure'' and larger than suspected structure
corrections. BBRW95 argue that this is not the case. The most massive
\hii\ regions in the sample are a diverse lot. The calculated
structure factors do allow for the possibility for ``microstructure''
below the angular resolution observed. The degree of such
microstructure is limited by observations of recombination lines.
Typically the calculated structure corrections are a few 10's\%. For
abundances consistent with chemical evolution models they would have
to be an order of magnitude larger.

 Another way the observed
correlation could arise is through local pollution. The trouble with
this scenario at first glance is that the stars which might plausibly
pollute \hii\ regions are massive, i.e., \he3 sinks.


It is generally agreed that \hii\ regions are ionized by massive stars
and that the most massive stars (O-stars eventually becoming
Wolf-Rayet stars) have very substantial winds which carry away most of
the stellar mass within their lifetimes. As far as we aware no
calculations have been published which give the \he3 abundance in
massive star winds.  However, it is plausible that the very earliest
winds are slightly enriched in \he3 from the initial ${\rm (D +
\he3)}$. From Maeder (1990) it seems possible that the first few
\msun\ of O-star/WR wind is \he3 rich. (The convective core
overshooting which contributes some uncertainty to abundances in WR
models [Schaller \etal\ 1992] will have no effect on the high \he3
material at the surface.) The later winds would be depleted in \he3
becoming first enhanced in N, then \he4, and finally C \& O. Thus, in
a young \hii\ region whose ionized gas was composed primarily by the
young winds of massive stars \he3 could be enhanced. Since the \he3
rich winds are a small fraction of the integrated wind mass loss, the
combined winds of many stars would be low in \he3 allowing even a
small dispersion in formation times. Only those regions containing a
very few (perhaps 1 or 2) stars would have high \he3.  W3, the \hii\
region with the highest observed \he3 could fit this model. W3A is a
bubble like structure with two embedded IR sources whose winds could
be shaping the region (Harris \& Wynn Williams 1976).  The region
observed by BBBRW (W3A plus some surrounding gas) is estimated to
contain about 15--25\msun\ of ionized gas. So a significant fraction
of the observed gas could be composed of slowed winds. W3 shows one
other sign of local pollution.  Roelfsema, Goss,
\& Mallik (1992) have observed substantial variations in the
\he4\ abundance in W3. Yet the overall \he4/H in W3 is ``normal''
(BBBRW). Our scenario suggests that winds in the W3 stars have just
reached the \he4 rich layers and that the \he4 rich blobs  are
slowed winds not yet mixed into the nebula as a whole.

As the evolution of an \hii\ region proceeds there are competing
factors which would determine the observed \he3 value. The later winds
would be very depleted in \he3, but some pristine gas from the ISM
containing some \he3 would be mixed in. If a \hii\ region were
composed almost entirely of late WR winds it could have a very low
\he3 but high \he4. Some limit on the admixture of wind gas and ISM
could be inferred from the observed \he4/H. It is curious that the
lowest \he3 abundance found is that in W49, the biggest \hii\ region
in the Galaxy which is estimated to contain many massive stars with a
total luminosity of $2 \times 10^6\,L_\odot$ (Dreher \etal\, 1984).
While it might be a candidate for substantial pollution by \he3 poor
winds, its ${\rm \he4/H} = 0.079$ does not suggest much pollution.

Note that any solution of this type, in which \he3 is depleted
by a rapid period of massive star formation will necessarily
predict an enhanced \he4  and heavy element abundances as discussed above.
However, as Lattimer, Schramm \& Grossman (1977) pointed out, the bulk
of the heavy element ejecta from supernovae can rapidly form into
dust grains.  These dust grains can behave like explosive ``shrapnel"
and penetrate regions exterior to the \hii\ region.  This would result
in the \hii\ region itself not showing a large heavy
element excess although the total heavy element enrichment would be part
of the integrated galactic enrichment. This is assuming of course
that the entire \hii\ region is not totally disrupted by the
supernovae explosion.

The \he3 data can be understood to be consistent with high primordial
D and \he3 abundances, \he3 production and galactic chemical
evolution, if one assumes that the \hii\ regions in which low \he3 is
observed are in fact biased tracers of the ISM \he3 abundance. Indeed,
\he3/H is lowest $\sim 10^{-5}$ in the most massive regions (of order a few
thousand solar masses) where there are many massive stars. If a
substantial part of the ionized gas is composed of stellar winds it
would be quite reasonable for these regions to be depleted in \he3.
Even the solar system could be depleted if the sun formed in early OB
association as has been suggested to account for various other
(heavier) isotopic anomalies (Olive \& Schramm 1981).  Any \hii\
region would be disrupted long before the low mass stars which produce
\he3 leave the main sequence.  However, it would appear that the only
way to lower the effective value of $g_3$ below that of the massive
stars (around 0.3) would be to argue that the gas in the region has
been cycled through stars several times.  Such an assumption however
would invariably predict \he4 abundances factors of 2--4 higher than
those observed.

Following this scenario only very young small \hii\ regions
10--20\msun\ which had been polluted by a few stars would show high
abundances of \he3. These \hii\ regions at their earliest stages could
provide a lower limit for the initial ${\rm D +~^3He}$ in the stars.

In conclusion, we have argued for the possibility that the
\he3 abundance in galactic \hii\ regions may be depleted and
therefore one should perhaps not compare directly results of chemical
evolution models with these abundances.  Similarly, solar system
abundances may be depleted if the solar system formed in an early OB
association.  While this is not a particularly palatable conclusion it
seems the best of the alternatives which we have considered.
In particular, the observations of high \he3 in planetary nebulae
clearly indicate that low mass stars must be net producers of \he3 in
agreement with calculations.  The \he3 observations are clearly of
great importance. Future observations of galactic \hii\
regions may also help in determining the degree of pollution in these
regions and the extent to which \he3 may be depleted.  We would
further argue that the apparent problems associated with \he3 are
therefore galactic rather than cosmological.  In that event, the
constraints on $\eta$ should remain intact.

\bigskip

\noindent {\bf Acknowledgments}

We would like to thank D. Balser, T. Bania, M. Cass\'{e}, B. Fields,
I. Iben, G. Steigman, F. Timmes, and T. Wilson for helpful
conversations.  The work of KAO was supported in part by DOE grant
DE-FG02-94ER-40823.  RTR was partially supported by NSF grant
AST--9121169. The work of DNS was supported in part by the DOE
(at Chicago and Fermilab) and by the NASA through NAGW-2381 (at
Fermilab) and a GSRP fellowship at Chicago.  The work of EV-F was
supported in part by PICS $n^o$114, ``Origin and Evolution of the
Light Elements", CNRS.

\newpage

\beginapjbib

\bibitem Balser, D.S., Bania, T.M., Brockway, C.J.,
Rood, R.T., \& Wilson, T.L. 1994, ApJ, 430, 667

\bibitem Balser, D.S., Bania, T.M.,
Rood, R.T., \& Wilson, T.L. 1995, in preparation

\bibitem Carswell, R.F., Rauch, M., Weymann, R.J., Cooke, A.J. \&
Webb, J.K. 1994, MNRAS, 268, L1

\bibitem Cass\'{e}, M. \& Vangioni-Flam, E. 1994,
talk presented at the ESO/EIPC Workshop on
the Light Element Abundances

\bibitem  Dearborn, D. S. P., Schramm, D.,
\& Steigman, G. 1986, ApJ, 302, 35

\bibitem Dreher, J. W., Johnston, K. J., Welch, W. J., \& Walker, R.
C. 1984, ApJ, 283,632

\bibitem Edvardsson, B., Anderson, J., Gustafsson, B.,
Lambert, D.L., Nissen, P.E., \& Tomkin, J. 1993,
A\&A, 275, 101

\bibitem Epstein, R., Lattimer, J., \& Schramm, D.N. 1976, Nature, 263, 198

\bibitem Galli, D., Palla, F., Straniero, O., \& Ferrini, F. 1994, ApJ,
432, L101

\bibitem Geiss, J. 1993, in {\it Origin
 and Evolution of the Elements} eds. N. Prantzos,
E. Vangioni-Flam, and M. Cass\'{e}
(Cambridge:Cambridge University Press), p. 89

\bibitem Harris, S., \& Wynn Williams, C. G. 1976, MNRAS, 174, 649

\bibitem Hartoog, M.R. 1979, ApJ, 231, 161

\bibitem Hobbs, L., \& Thorburn, J., 1994, ApJ, 428, L25

\bibitem Hogan, C.J. 1994, University of Washington preprint

\bibitem Iben, I. 1967, ApJ, 147, 624

\bibitem Iben, I. \& Truran, J.W. 1978, ApJ, 220,980

\bibitem Iben, I. \& Tutukov, A. 1984, ApJ Supp, 54, 335

\bibitem Lattimer, J., Schramm, D.N., \& Grossman, L. 1977, ApJ, 214, 819

\bibitem Linsky, J.L., \etal\ 1992, ApJ, 402, 694

\bibitem Linsky, J.L. 1994,  talk presented at the ESO/EIPC Workshop on
the Light Element Abundances

\bibitem Maeder, A. 1990, A \& A Supp, 84, 139

\bibitem Olive, K.A., \& Schramm, D.N. 1981, ApJ, 257, 276

\bibitem Olive, K.A., \& Steigman, G. 1994, University of Minnesota preprint
UMN-TH-1230/94.

\bibitem Ostriker, J.P., \& Schramm, D.N. 1994, in preparation

\bibitem Pagel, B E.J., Simonson, E.A., Terlevich, R.J.
\& Edmunds, M. 1992, MNRAS, 255, 325

\bibitem Renzini, A. \& Voli, M. 1981, A\&A, 94, 175

\bibitem Roelfsema, P. R., Goss, W. M., \& Mallik, D. C. V. 1992, ApJ,
394, 188

\bibitem Rood, R.T. 1972, ApJ, 177, 681

\bibitem Rood, R.T., Bania, T.M., \& Wilson, T.L. 1984, ApJ, 280, 629

\bibitem Rood, R.T., Bania, T.M., \& Wilson, T.L. 1992, Nature, 355, 618

\bibitem  Rood, R. T., Bania, T. M., Wilson, T. L., \& Balser, D. S. 1995,
in ESO/EPIC Workshop: The Light Elements, ed. P. Crane (Springer: Berlin)

\bibitem Rood, R.T., Steigman, G. \& Tinsley, B.M. 1976, ApJ, 207, L57

\bibitem Schaller, G., Schaerer, D., Meynet, G., \& Maeder, A. 1992,
A\&AS, 96, 269

\bibitem Skillman, E., \etal\ 1994b, ApJ Lett (in preparation)

\bibitem Smith, V.V., Lambert, D.L., \& Nissen, P.E., 1992, ApJ 408, 262

\bibitem Songaila, A., Cowie, L.L., Hogan, C. \& Rugers, M. 1994
Nature, 368, 599

\bibitem Spite, F. \& Spite, M. 1993, in {\it Origin
 and Evolution of the Elements} eds. N. Prantzos,
E. Vangioni-Flam, and M. Cass\'{e}
(Cambridge:Cambridge University Press), p.201

\bibitem Steigman, G. 1994, Ohio State University preprint OSU-TA-7/94

\bibitem Steigman, G., Fields, B. D., Olive, K. A., Schramm, D. N.,
\& Walker, T. P., 1993, ApJ 415, L35

\bibitem Steigman, G. \& Tosi, M. 1992, ApJ, 401, 150

\bibitem Thorburn, J. 1993, ApJ, 421, 318

\bibitem Tosi, M. 1988, A\&A, 197, 33

\bibitem Tosi, M. 1994,  talk presented at the ESO/EIPC Workshop on
the Light Element Abundances

\bibitem Vangioni-Flam, E. \& Cass\'{e}, M. 1994, ApJ (submitted)

\bibitem Vangioni-Flam, E., Olive, K.A., \& Prantzos, N. 1994,
ApJ, 427, 618

\bibitem Vassiliadis, E. \& Wood, P.R. 1993, ApJ, 413, 641

\bibitem Walker, T. P., Steigman, G., Schramm, D. N., Olive, K. A.,
\& Kang, H. 1991 ApJ, 376, 51

\bibitem Yang, J., Turner, M.S., Steigman, G., Schramm, D.N., \& Olive, K.A.
1984,
ApJ, 281, 493.

\endapjbib

\newpage
\begin{table}[h]
\begin{center} {\sc Model Results}:
\end{center}
 $\sigma_0$
denotes the present gas mass fraction, ${\rm (D/H)}_\odot$ the
protosolar value of the deuterium to hydrogen ratio, and the
associated destruction factors ${\rm D}_p/{\rm D}_\odot$ and
${\rm D}_p/{\rm D}_0$ are evaluated at solar birth and at the present,
$Z$ is the overall metallicity.  Models 1,2,3 differ by the primordial D/H
abundance and hence the adopted value of $g_3$ and the SFR required
to obtain the present D/H value.  Model 2.1 is similar to model 2
except for the chosen value of $g_3$ (see text).
\begin{center}
\begin{tabular}{lccccc}                                           \hline \hline
             & Observations & Model 1 & Model 2 & Model 2.1 & Model 3  \\
\hline
$\sigma_0$     & 0.1 to 0.2  &   0.13 & 0.17 & 0.17  & 0.18   \\
${\rm (D/H)}_\odot$  &$(2.6 \pm 1.0) \times 10^{-5}$ & $3.3 \times
10^{-5}$&$3.2 \times 10^{-5}$ &
$3.2 \times 10^{-5}$ & $1.9 \times 10^{-5}$ \\
${\rm D}_p/{\rm D}_\odot$ & &2.3&7.8&   7.8    &   1.8     \\
${\rm D}_p/{\rm D}_0$  & &4.3&12&12& 3  \\
${\rm (\he3/H)}_\odot$ &$(1.5 \pm 0.3) \times 10^{-5}$&$5.2  \times 10^{-5}$
    &$1.8  \times 10^{-4}$&$1.9  \times 10^{-5}$&$3.4  \times 10^{-5}$   \\
$\left({\rm \frac{(D+~^3He)}{H}}\right)_\odot$  &$(4.1 \pm 1.0) \times 10^{-5}$
&$8.5 \times 10^{-5}$&
  $2.1 \times 10^{-4}$&  $5.1 \times 10^{-5}$& $5.3 \times 10^{-5}$  \\
$Z/Z_\odot$ &1&1.7&2.1&2.1&1.4 \\ \hline
\end{tabular}

\end{center}
\end{table}
\newpage
\noindent{\bf{Figure Captions}}

\vskip.3truein

\begin{itemize}

\item[]
\begin{enumerate}
\item[]
\begin{enumerate}

\item[{\bf Figure 1:}] The differential yield of \he3 as a
function of stellar mass. The \he3 yield is taken from eq. (\ref{it})
and from Dearborn, Schramm, \& Steigman (1986).
Other parameters are those from Model 1.

\item[{\bf Figure 2:}]  The evolution of
D/H (dashed curve), \he3/H (solid curve) and (D + \he3)/H (dotted curve)
  as a function of time.  Also shown are the data at the solar epoch
$t \approx 9.6$ Gyr and today for
D/H (open squares), \he3/H (filled diamonds) and (D + \he3)/H (open circle).
The chemical evolution model has been chosen so that D/H agrees with the data.
The problem we are emphasizing is with \he3 and can be seen by comparing
the solid curve with the filled diamonds. A primordial value
of D/H $= 7.5 \times 10^{-5}$ was chosen.

\item[{\bf Figure 3:}] As in Figure 2, with a primordial value
of D/H $= 2.5 \times 10^{-4}$.

\item[{\bf Figure 4:}] As in Figure 2, with a primordial value
of D/H $= 3.5 \times 10^{-5}$.

\item[{\bf Figure 5:}] The \he3/H abundance in several galactic \hii\
regions as a function of the mass of the region (from BBRW95).

\end{enumerate}
\end{enumerate}
\end{itemize}

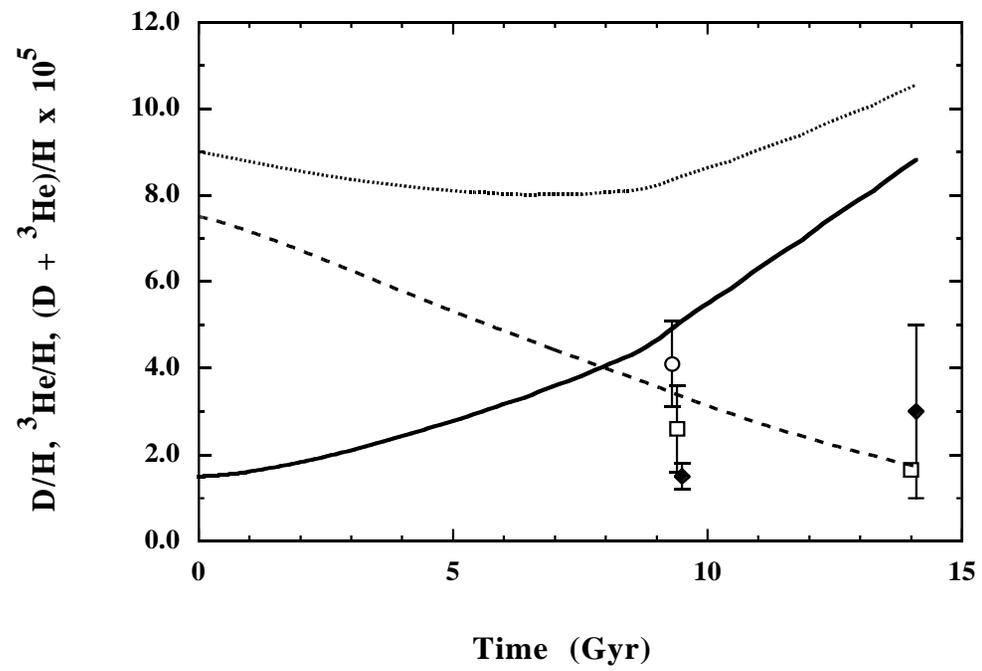
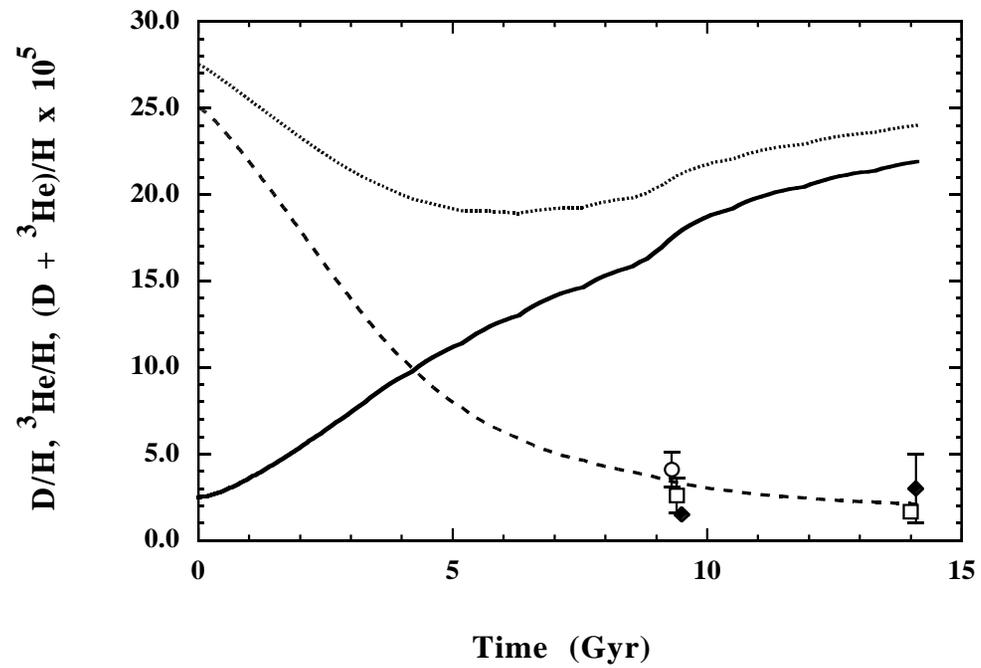
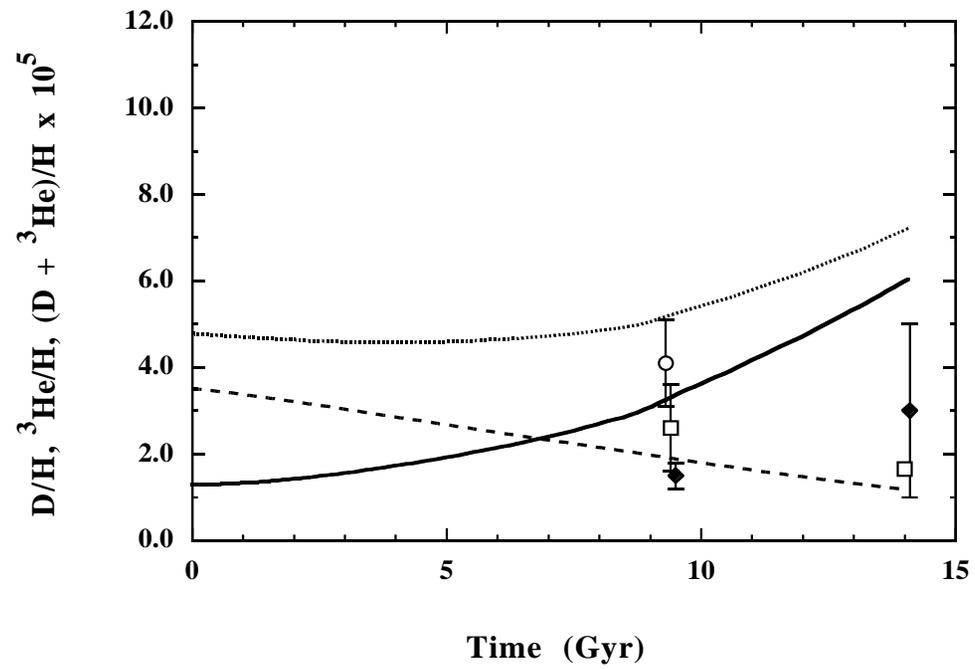
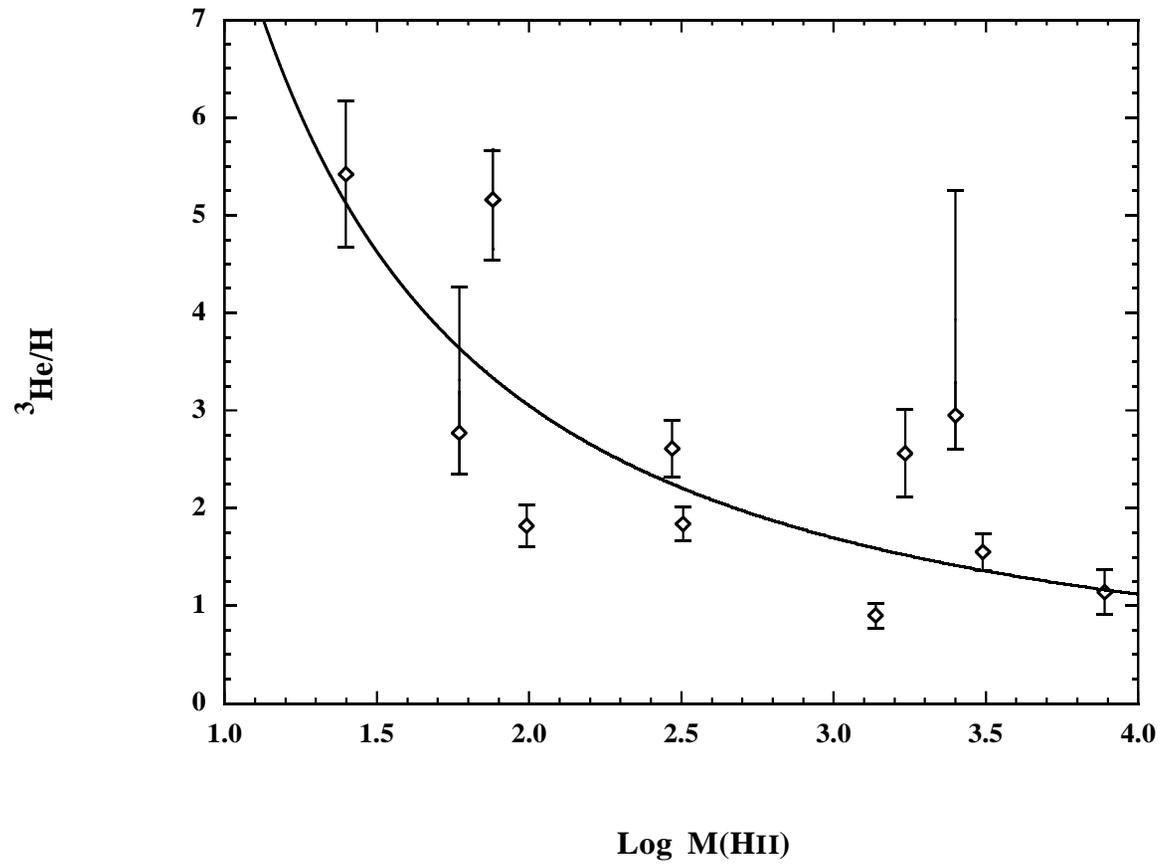
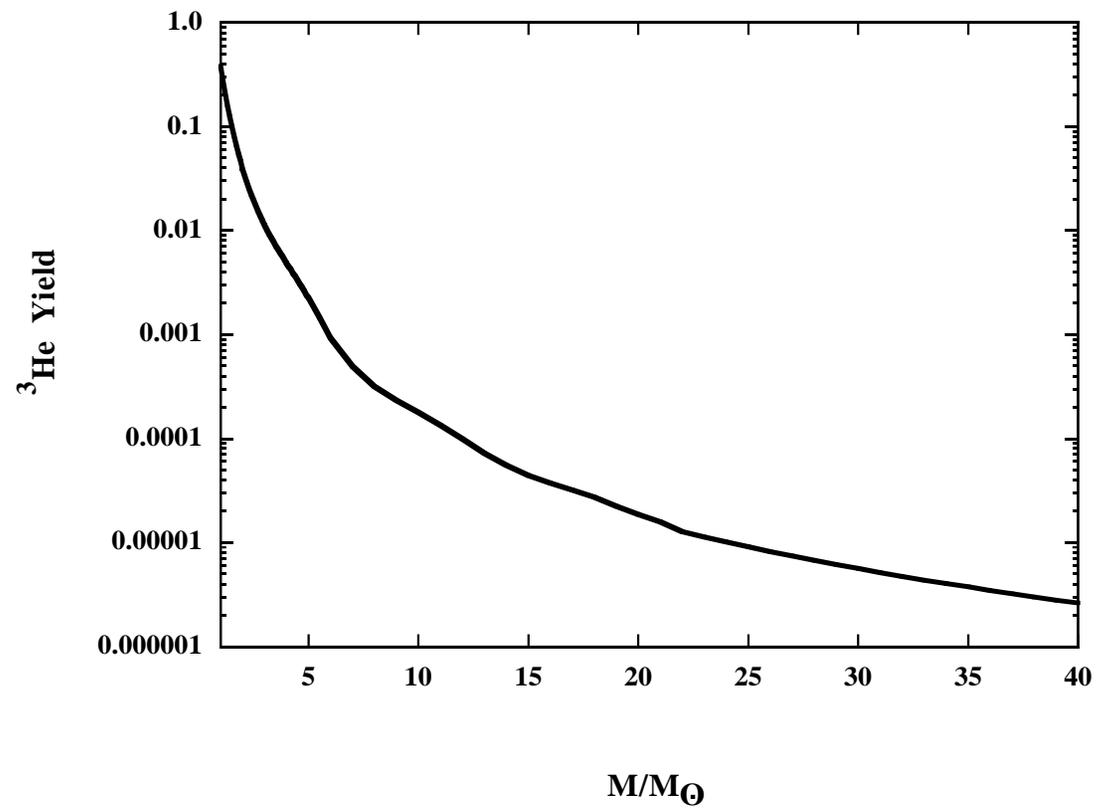
\end{document}